\documentclass[twocolumn,amsmath,amssymb,letter]{revtex4}
\usepackage{epsfig}
\usepackage{bm}
\usepackage{graphicx}
\usepackage{dcolumn}% Align table columns on decimal point

\begin{document}

\title{Nonlocal Form of the Rapid Pressure-Strain Correlation
in Turbulent Flows}
\author{Peter E. Hamlington
\footnote{Corresponding author, email: peterha@umich.edu}}
\affiliation{Laboratory for Turbulence \& Combustion (LTC),
Department of Aerospace Engineering, The University of Michigan,
Ann Arbor, MI 48109-2140, USA}

\author{Werner J.A. Dahm}
\affiliation{Laboratory for Turbulence \& Combustion (LTC),
Department of Aerospace Engineering, The University of Michigan,
Ann Arbor, MI 48109-2140, USA}

\date{\today}

\begin{abstract}

A new fundamentally-based formulation of nonlocal effects in the
rapid pressure-strain correlation in turbulent flows has been
obtained. The resulting explicit form for the rapid
pressure-strain correlation accounts for nonlocal effects produced
by spatial variations in the mean-flow velocity gradients, and is
derived through Taylor expansion of the mean velocity gradients
appearing in the exact integral relation for the rapid
pressure-strain correlation. The integrals in the resulting series
expansion are solved for high- and low-Reynolds number forms of
the longitudinal correlation function $f(r)$, and the resulting
nonlocal rapid pressure-strain correlation is expressed as an
infinite series in terms of Laplacians of the mean strain rate
tensor. The new formulation is used to obtain a nonlocal transport
equation for the turbulence anisotropy that is expected to provide
improved predictions of the anisotropy in strongly inhomogeneous
flows.

\end{abstract} \noindent
%\pacs{47.27.ek, 83.10.Mj, 83.80.Rs}

\maketitle
\pagebreak

\section{Introduction}

By far the most practical approaches for simulating turbulent
flows are based on the ensemble-averaged Navier-Stokes equations.
However, such approaches require a suitably accurate closure model
for the Reynolds stress anisotropy tensor $a_{ij}$, defined as
\begin{equation}\label{rans5}
   a_{ij} \equiv \frac{\overline{u'_i
u'_j}}{k} - \frac{2}{3} \delta_{ij}\,,
\end{equation}
where $\overline{u'_i u'_j}$ are the Reynolds stresses and $k
\equiv \frac{1}{2} \overline{u'_i u'_i}$ is the turbulence kinetic
energy. Over the past half century, a wide range of closures for
$a_{ij}$ have been proposed. Of these, so-called Reynolds stress
transport models that solve the full set of coupled partial
differential equations for $a_{ij}$ are currently regarded as
having the highest fidelity among practical closures. Such
closures start from the exact transport equation for $a_{ij}$,
namely
\begin{eqnarray}\label{es4}
  \frac{Da_{ij}}{Dt} = -\left[\frac{P}{\epsilon} - 1\right]
  \frac{\epsilon}{k} a_{ij} + \frac{1}{k} \left[ P_{ij} -
  \frac{2}{3} P \delta_{ij}\right]+ \frac{1}{k} \Pi_{ij}
  \hspace{0in}\\ \hspace{0in}-\frac{1}{k}
  \left[\epsilon_{ij} - \frac{2}{3}\epsilon \delta_{ij}\right]+
  \frac{1}{k} \left[D_{ij} - \left(a_{ij} +
  \frac{2}{3}\delta_{ij}\right) D\right]\,,\nonumber
\end{eqnarray}
where for clarity we are restricting the presentation to
incompressible flows.  In (\ref{es4}), $D/Dt$ is the mean-flow
material derivative, $P_{ij}\equiv -(\overline{u'_i u'_l}\
\partial \overline{u}_j /\partial x_l +\overline{u'_j u'_l}\
\partial \overline{u}_i /\partial x_l)$ is the production tensor,
$\epsilon_{ij}$ is the dissipation tensor, and all remaining
viscous, turbulent, and pressure transport terms are contained in
$D_{ij}$, with $P\equiv P_{nn}/2$, $\epsilon \equiv
\epsilon_{nn}/2$, and $D \equiv D_{nn}/2$. In such Reynolds stress
transport closures, the production tensor needs no modeling since
$\overline{u'_i u'_j}$ is obtained from $a_{ij}$, and standard
models for $\epsilon_{ij}$ and $D_{ij}$ are discussed in Refs.\
\cite{speziale1991,speziale1998}. The principal remaining
difficulty is in accurately representing $\Pi_{ij}$ in
(\ref{es4}), namely the pressure-strain correlation tensor
\begin{equation}\label{pstr1}
  \Pi_{ij}(\textbf{x}) \equiv \frac{2}{\rho}\, \overline{p'(\textbf{x})
   S'_{ij}(\textbf{x})}\,,
\end{equation}
where
\begin{equation}\label{pstr2}
  S'_{ij} \equiv \frac{1}{2} \left( \frac{\partial u'_i }{\partial x_j}
  + \frac{\partial u'_j} {\partial x_i}\right)
\end{equation}
are the strain rate fluctuations. The pressure-strain correlation
has received considerable attention, however developing a
fundamentally-based yet practically implementable form for
$\Pi_{ij}$ remains one of the primary challenges in turbulence
research.

The difficulty in representing $\Pi_{ij}$ stems in large part from
the inherently nonlocal nature of the pressure-strain correlation,
since the local pressure $p'$ in (\ref{pstr1}) depends on an
integral over the entire spatial domain of the flow. Some progress
has been made by splitting $\Pi_{ij}$ into the sum of ``slow'' and
``rapid'' parts \cite{chou1945}, where the rapid part
$\Pi^{(r)}_{ij}$ is so named due to its direct dependence on the
mean-flow velocity gradients $\partial \overline{u}_i / \partial
x_j$, variations in which have an immediate effect on
$Da_{ij}/Dt$. Typically, the slow part $\Pi^{(s)}_{ij}$ is
represented in terms of the local values of $a_{ij}$ and
$\epsilon$. For the rapid part, it has been common (\textit{e.g.,}
\cite{chou1945,rotta1951,crow1968}) to take the mean velocity
gradients as being sufficiently homogeneous that they can be
brought outside the integral. Under certain conditions
\cite{crow1968} the remaining integral can then be solved for the
local part of $\Pi^{(r)}_{ij}$.  This is then typically combined
with additional \textit{ad hoc} terms involving $a_{ij}$ to model
the rapid part solely in terms of local flow variables. Together
with the assumed local representation for the slow part, this
yields a purely local formulation for $\Pi_{ij}$ that allows
(\ref{es4}) to be solved, but that neglects all nonlocal effects
in the evolution of the anisotropy.

Such purely local models for $\Pi_{ij}$ have allowed relatively
accurate simulations of homogeneous turbulent flows, where by
construction there are no spatial variations in $\partial
\overline{u}_i / \partial x_j$ and thereby all nonlocal effects
vanish.  However most practical situations involve strongly
inhomogeneous flows, where large-scale structure and other
manifestations of spatial variations in the mean-flow velocity
gradients can produce significant nonlocal effects in the
turbulence, the neglect of which in $\Pi_{ij}$ can lead to
substantial inaccuracies in the resulting anisotropy. Such
nonlocal effects are significant even in free shear flows such as
jets, wakes, and mixing layers, and can become especially
important in near-wall flows, where flow properties vary rapidly
in the wall-normal direction. Improving the fidelity of turbulent
flow simulations requires a fundamentally-based formulation for
nonlocal effects in $\Pi^{(r)}_{ij}$ to account for spatial
variations of velocity gradients in the ensemble-averaged flow.

Various methods for addressing such spatial variations have been
proposed, however nearly all suffer from a lack of systematic
physical and mathematical justification. For near-wall flows, by
far the most common yet also least satisfying approach is the use
of empirical ``wall damping functions" (\textit{e.g.},
\cite{speziale1998}). Although such functions are relatively
straightforward to implement, they are also distinctly \textit{ad
hoc} and as a consequence do not perform well across a wide range
of flows. Moreover, wall functions typically conflate the
treatment of a number of near-wall effects that in fact originate
from distinctly different physical mechanisms, including low
Reynolds number effects, large strain effects, and wall-induced
kinematic effects, and are not formulated to specifically account
for nonlocality due to spatial variations in the mean flow
gradients.

In the following we depart from these prior approaches by
systematically deriving a new nonlocal formulation for the rapid
pressure-strain correlation from the exact integral relation for
the rapid part of $\Pi_{ij}$. Specifically, nonlocal effects due
to mean-flow velocity gradients are accounted for through Taylor
expansion of $\partial \overline{u}_k / \partial x_l$ in the rapid
pressure-strain integral. The resulting nonlocal form of the rapid
pressure-strain correlation $\Pi^{(r)}_{ij}$ appears as a series
of Laplacians of the mean strain rate tensor. The only
approximation involved -- beyond the central hypothesis on which
the present formulation is based -- is an explicit form for the
longitudinal correlation function $f(r)$, though the effect of
this is only to determine specific values of the coefficients in
an otherwise fundamental result for the nonlocal effects in
$\Pi^{(r)}_{ij}$. The coefficients are obtained here for the
exponential form of $f(r)$ appropriate for high Reynolds numbers,
and for the exact Gaussian $f(r)$ that applies at low Reynolds
numbers. The resulting formulation for the rapid part of
$\Pi_{ij}$ then provides a new nonlocal anisotropy transport
equation that can be used with any number of closure approaches
for representing $a_{ij}$, including Reynolds stress transport
models as well as explicit stress models suitable for two-equation
closures.

\section{Nonlocal Formulation for the Pressure-Strain Correlation}

The starting point for developing a fundamentally-based
representation for $\Pi_{ij}$ is the exact Poisson equation
for the pressure fluctuations
$p'$ appearing in (\ref{pstr1}), namely
\begin{equation}\label{ps1}
  \frac{1}{\rho} \nabla^2 p' = -2 \frac{\partial
  \overline{u}_k}{\partial x_l}\frac{\partial u'_l}{\partial x_k}
  - \frac{\partial^2}{\partial x_k
  \partial x_l} \left( u'_k u'_l - \overline{u'_k u'_l}\right)\,
\end{equation}
(\textit{e.g.,} \cite{pope2000}).
Beginning with Chou \cite{chou1945}, it has been common
to write $p'$ in terms of rapid, slow, and wall parts as
\begin{equation}\label{ps2.1}
  p' \equiv p'^{(r)}+p'^{(s)}+p'^{(w)}\,,
\end{equation}
defined by their respective Poisson equations from
(\ref{ps1}) as
\begin{equation}\label{ps2.3}
  \frac{1}{\rho}\nabla^2 p'^{(r)} =  -2 \frac{\partial
  \overline{u}_k}{\partial x_l}\frac{\partial u'_l}{\partial
  x_k}\,,
\end{equation}
\begin{equation}\label{ps2.2}
  \frac{1}{\rho}\nabla^2 p'^{(s)} =  - \frac{\partial^2}{\partial
  x_k
  \partial x_l} \left( u'_k u'_l - \overline{u'_k u'_l}\right)\,,
\end{equation}
\begin{equation}\label{ps2.4}
  \frac{1}{\rho} \nabla^2 p'^{(w)}=0\,.
\end{equation}
The effect of $p'^{(w)}$ is significant in (\ref{es4}) only in the
extreme near-wall region of wall-bounded flows
\cite{pope2000,mansour1988}. The remaining rapid and slow parts
produce corresponding rapid and
slow contributions to the pressure-strain correlation $\Pi_{ij}$
in (\ref{pstr1}), with Green's function solutions of (\ref{ps2.3})
and (\ref{ps2.2}) giving these as
\begin{equation}\label{ps4.1}
  \Pi^{(r)}_{ij}(\textbf{x})=\frac{1}{\pi} \int_{\bf R}
  \frac{\partial
  \overline{u}_k(\hat{\textbf{x}})}{\partial \hat{x}_l}
  \overline{\frac{\partial u'_l(\hat{\textbf{x}})}{\partial
  \hat{x}_k}S'_{ij}(\textbf{x})}
  \frac{d^3 \hat{\textbf{x}}}{|\textbf{x}-\hat{\textbf{x}}|}\,
\end{equation}
\begin{equation}\label{ps4.2}
  \Pi^{(s)}_{ij}(\textbf{x})=\frac{1}{2\pi} \int_{\bf R}
  \overline{\frac{\partial^2
  \left( u'_k u'_l\right)_{\hat{\textbf{x}}}}{\partial
  \hat{x}_k  \partial \hat{x}_l}
  S'_{ij}(\textbf{x})}
  \frac{d^3 \hat{\textbf{x}}}{|\textbf{x}-\hat{\textbf{x}}|}\,,
\end{equation}
where the integration spans the entire flow domain $\textbf{R}$.

The slow part $\Pi^{(s)}_{ij}$ is typically not treated in a
systematic fashion via integration of (\ref{ps4.2}). Instead,
nearly all existing representations for $\Pi_{ij}^{(s)}$ are based
on insights obtained from the return to isotropy of various forms
of initially-strained grid turbulence. The most common
representation for $\Pi_{ij}^{(s)}$ is Rotta's \cite{rotta1951}
linear ``return-to-isotropy'' form
\begin{equation}\label{slow1}
  \Pi_{ij}^{(s)} = -C_1 \epsilon a_{ij}\,,
\end{equation}
where all variables are local and $C_1$ is typically in the range
$1.5-1.8$
(\textit{e.g.,} \cite{speziale1998,pope2000}). Sarkar and Speziale
\cite{sarkar1990,speziale1991a} have argued that additional
quadratic terms should be included in (\ref{slow1}), but it has
been noted \cite{speziale1998} that these are typically small. As
a result, representations for $\Pi_{ij}^{(s)}$ remain relatively
simple, and the form in (\ref{slow1}) continues to be widely used.

By contrast, $\Pi_{ij}^{(r)}$ has received substantially greater
attention. The direct effect of the mean velocity gradients
$\partial\overline{u}_k / \partial x_l$ on this rapid part of the
pressure-strain correlation is apparent in (\ref{ps4.1}).  In the
following sections, we use the integral in (\ref{ps4.1}) to
develop a fundamentally-based representation for $\Pi_{ij}^{(r)}$
that accounts for nonlocal effects resulting from spatial
nonuniformities in the mean velocity gradients.

\subsection{Prior Local Formulation for $\Pi_{ij}^{(r)}(\textbf{x})$}
Chou \cite{chou1945} first suggested the notion of using the
integral form in (\ref{ps4.1}) to obtain a representation for the
rapid pressure-strain correlation. Subsequently, Rotta
\cite{rotta1951} and then Crow \cite{crow1968} used that approach
to rigorously derive the purely local part of $\Pi_{ij}^{(r)}$, by
assuming the mean velocity gradients in (\ref{ps4.1}) to vary
sufficiently slowly that they could be taken as constant over the
length scale on which the two-point correlation
$\overline{\left[\partial
u'_k(\hat{\textbf{x}})/\partial\hat{x}_l\right]
S'_{ij}(\textbf{x})}$ in (\ref{ps4.1}) is nonzero. Under such
conditions, the mean velocity gradient in (\ref{ps4.1}) can be
taken outside the integral, and $\Pi_{ij}^{(r)}$ then becomes
\begin{equation}\label{newrapidform1}
  \Pi^{(r)}_{ij}(\textbf{x}) \approx
  \frac{\partial\overline{u}_k(\textbf{x})}{\partial {x}_l}
  \cdot
  \frac{1}{\pi} \int_{\bf R}
  \overline{\frac{\partial u'_l(\hat{\textbf{x}})}{\partial
  \hat{x}_k}S'_{ij}(\textbf{x})}
  \frac{d^3 \hat{\textbf{x}}}{|\textbf{x}-\hat{\textbf{x}}|}\,.
\end{equation}
With $S'_{ij}(\bf{x})$ in (\ref{pstr1}), the integrand in
(\ref{newrapidform1}) involves two-point correlations among
velocity gradients of the form
\begin{equation}\label{twopointcorr}
  \overline{\frac{\partial u'_i(\textbf{x})}{\partial x_j}
  \frac{\partial u'_l(\hat{\textbf{x}})}{\partial \hat{x}_k}}
   = -\frac{\partial^2 R_{il}(\textbf{r})}{\partial r_j\partial
  r_k}\,,
\end{equation}
where $R_{il}(\textbf{r})$ denotes the velocity fluctuation
correlation
\begin{equation}\label{Rijdefinition}
R_{il}(\textbf{r})\equiv \overline{u'_i
  (\textbf{x}) u'_l(\textbf{r}+\textbf{x})}
  \end{equation}
with $\textbf{r}\equiv\hat{\textbf{x}}-\textbf{x}$.
Defining \cite{chou1945,rotta1951,crow1968}
\begin{equation}\label{ps6}
  M_{iljk} \equiv -\frac{1}{2\pi} \int_{\bf R}
  \frac{\partial^2 R_{il}(\textbf{r})}{\partial r_j\partial r_k}
  \frac{d^3\textbf{r}}{r}\,,
\end{equation}
the rapid pressure-strain correlation in (\ref{newrapidform1}) can
then be expressed as
\begin{equation}\label{ps5}
  \Pi_{ij}^{(r)}(\textbf{x}) \approx \frac{\partial
  \overline{u}_k(\textbf{x})}{\partial x_l}\left[
  M_{iljk}+
  M_{jlik}\right]\,.
\end{equation}

Using the homogeneous isotropic form of $R_{il}(\textbf{r})$,
namely
\begin{equation}\label{Rijform}
  R_{il}(r) = \frac{2}{3}k
   \left[ f(r)\delta_{il}+
  \frac{r}{2} \frac{df}{dr}
  \left( \delta_{il} - \frac{r_i
  r_l}{r^2}\right)\right]\,
\end{equation}
with
\begin{equation}\label{frdef}
  f(r) \equiv \frac{3}{2}\frac{\overline{u'(\textbf{x}+\textbf{r})
  u'(\textbf{x})}}{k}\,,
\end{equation}
where $k$ is the turbulence kinetic energy, it can be shown
\cite{crow1968} that $M_{iljk}$ in (\ref{ps6}) becomes
\begin{equation}\label{ps7}
  M_{iljk} = \frac{2}{15} k \left(4\delta_{jk} \delta_{il} -
  \delta_{ij} \delta_{kl} - \delta_{jl}\delta_{ki}\right)\,,
\end{equation}
where the leading $k$ again denotes the turbulence kinetic
energy.
Using (\ref{ps7}) in (\ref{ps5}) then gives the rapid pressure-strain
correlation as
\begin{equation}\label{ps8}
  \frac{1}{k} \Pi_{ij}^{(r)} \approx \frac{4}{5} \overline{S}_{ij}\,,
\end{equation}
where $\overline{S}_{ij}$ is the local mean-flow strain rate tensor
\begin{equation}
  \overline{S}_{ij} \equiv \frac{1}{2}
  \left( \frac{\partial \overline{u}_i }{\partial x_j}
  + \frac{\partial \overline{u}_j} {\partial
  x_i}\right)\,.
\end{equation}

Typically, (\ref{ps8}) is used as the leading-order isotropic term
in tensorial expansions for the rapid pressure-strain correlation
$\Pi_{ij}^{(r)}(\textbf{x})$, where the remaining terms are
expressed in terms of the local anisotropy $a_{ij}$ and the local
mean velocity gradient tensor.
Note however that such representations are still purely local,
since in going from (\ref{ps4.1}) to (\ref{newrapidform1}) all
spatial variations in the mean velocity gradients
$\partial\overline{u}_k / \partial x_l$ over the length scale on
which the two-point correlations $\overline{\left[\partial
u'_k(\hat{\textbf{x}})/\partial\hat{x}_l\right]
S'_{ij}(\textbf{x})}$ are nonzero were ignored. The resulting
neglect of nonlocal contributions to $\Pi^{(r)}_{ij}$ from that
approximation can lead to substantial inaccuracies in many
turbulent flows, including free shear flows and wall-bounded
flows. For example, Bradshaw \textit{et al.} \cite{bradshaw1987}
showed using DNS of fully-developed turbulent channel flow
\cite{kim1987} that the homogeneity approximation used to obtain
(\ref{ps5}) is invalid for $y^+\leq 30$. It can be further shown
(\textit{e.g.}, \cite{kim1987,iwamoto2002}) that  the dominant
component $\overline{S}_{12}$ of the mean strain
begins to vary dramatically at locations as far from the wall as
$y^+\approx 60$. Comparable variations in mean velocity gradients
are also found in turbulent jets, wakes, and mixing layers, where
there are substantial spatial variations in $\overline{S}_{12}$
across the flow. Indeed in most turbulent flows of practical
interest, there are significant variations in the mean flow
velocity gradients that will produce nonlocal contributions to the
rapid pressure-strain correlation via (\ref{ps4.1}). In such
situations, it may be essential to account for these nonlocal
effects in $\Pi_{ij}^{(r)}$ to obtain accurate results from any
closures based on (\ref{es4}).

\subsection{Present Nonlocal Formulation for $\Pi_{ij}^{(r)}(\textbf{x})$}
\label{derive}

In the following, nonlocal effects due to spatial variations in
the mean flow are accounted for in $\Pi_{ij}^{(r)}$ through Taylor
expansion of the mean velocity gradients appearing in
(\ref{ps4.1}). The central hypothesis in the approach developed
here is that the nonlocality in $\Pi^{(r)}_{ij}$ is substantially due
to spatial variations in
$\partial\overline{u}_k / \partial x_l$ in (\ref{ps4.1}),
and that in order to address this effect all
other factors in (\ref{ps4.1}) can be adequately represented by
their homogeneous isotropic forms. This allows a formulation of
the rapid pressure-strain correlation analogous to that in
(\ref{ps8}), but goes beyond a purely local formulation to take
into account the effects of spatial variations in the mean flow
gradients.

We begin by defining the ensemble-averaged velocity gradients
\begin{equation}\label{Aij}
A_{kl} \equiv \partial \overline{u}_k/\partial x_l\,,
\end{equation}
and account for spatial variations in $A_{kl}(\hat{\textbf{x}})$
in (\ref{ps4.1}) via its local Taylor expansion about the point
$\textbf{x}$ as
\begin{eqnarray}\label{gen0}
  A_{kl}(\hat{\textbf{x}})=
  A_{kl}(\textbf{x})+ r_m \frac{\partial
  A_{kl}}
  {\partial x_m}+ \frac{r_m r_p}{2}
  \frac{\partial^2 A_{kl}}
  {\partial x_m \partial x_p}\\
  +\cdots + \frac{1}{n!} \left( r_m r_p
  \ldots\right) \frac{\partial^n A_{kl}}{\partial x_m
  \partial x_p\ldots}\,,\nonumber
  \end{eqnarray}
where $\textbf{r} \equiv \hat{\textbf{x}}-\textbf{x}$ and all
derivatives of $A_{kl}$ are evaluated at $\textbf{x}$, and where
$n$ is the order of the expansion. As $n\rightarrow \infty$, the
expansion provides an exact representation of all spatial
variations in $A_{kl}(\textbf{r}+\textbf{x})$ from purely local
information at \textbf{x}. Substituting (\ref{gen0}) into
(\ref{ps4.1}) then gives
\begin{equation}\label{gen1}
  \Pi_{ij}^{(r)}(\textbf{x}) = \sum_{n=0}^\infty
  \frac{\partial^n A_{kl}(\textbf{x})}
  {\partial x_m \partial x_p\ldots}
   \left[_{(mp\ldots)}M^{(n)}_{iljk}
   +_{(mp\ldots)}M^{(n)}_{jlik}\right]\,,
\end{equation}
where
\begin{equation}\label{gen1.1}
  _{(mp\ldots)}M^{(n)}_{iljk}\equiv
  -\frac{1}{2\pi n!} \int_{\bf R}
  \left[ \frac{ r_m r_p
  \ldots}{r^n}\right]r^{n-1} \frac{\partial^2
R_{il}(\textbf{r})}{\partial
  r_j r_k} d^3 \textbf{r}\,.
\end{equation}
The $n$th-order term in (\ref{gen1}) involves $n$
derivatives of $A_{kl}$ as well as $n$ total indices
$(mp\ldots)$ in $_{(mp\ldots)}M^{(n)}_{iljk}$.

From the central hypothesis on which the present treatment of
nonlocal effects in $\Pi^{(r)}_{ij}$ is based, we represent
$R_{il}(\textbf{r})$ in (\ref{gen1.1}) by the form in
(\ref{Rijform}). With the relations
\begin{equation}\label{pope16}
  \frac{\partial r}{\partial r_j} = \frac{r_j}{r},\quad
  \frac{\partial r_i}{\partial r_j} = \delta_{ij}\,,
\end{equation}
the double derivative of $R_{il}(\textbf{r})$ in (\ref{gen1.1}) is
then given by
\begin{equation}\label{pope18}
  \frac{\partial^2 R_{il}(\textbf{r})}{\partial r_j \partial r_k}
  = \frac{k}{3} \left[a_{ijkl}\frac{1}{r} \frac{df}{dr}  +
  b_{ijkl} \frac{d^2 f}{dr^2}  + c_{ijkl} r\frac{d^3f}{dr^3}
  \right]\,,
\end{equation}
where we have introduced the compact notation
\begin{subequations}\label{pope19}
\begin{eqnarray}
  a_{ijkl} \equiv 3\delta_{jk}\delta_{il} - \delta_{ij} \delta_{kl} -
  \delta_{jl}\delta_{ki}- 3\alpha_{jk} \delta_{il} + \delta_{ij}
  \alpha_{lk} \\+ \delta_{jl}\alpha_{ik} + \delta_{ik} \alpha_{lj}
  +  \delta_{kl} \alpha_{ij} + \delta_{kj} \alpha_{il} - 3
  \beta_{iljk}\,,\nonumber
\end{eqnarray}
\begin{eqnarray}
  b_{ijkl} \equiv \delta_{il}\delta_{jk} + 3\alpha_{jk}\delta_{il} -
  \delta_{ij}\alpha_{lk} - \delta_{lj}\alpha_{ik} - \delta_{ik}
  \alpha_{lj}\\ - \delta_{kl}\alpha_{ij} - \delta_{kj}\alpha_{il} +
  3\beta_{iljk}\,,\nonumber
\end{eqnarray}
\begin{equation}
  c_{ijkl} \equiv \delta_{il}\alpha_{jk} - \beta_{iljk}\,,
\end{equation}
\end{subequations}
with
\begin{equation}\label{pope22}
\alpha_{ij} \equiv \frac{r_i r_j}{r^2}\,, \quad
  \beta_{ijkl} \equiv \frac{r_i r_j r_k r_l}{r^4}\,.
\end{equation}
Writing the differential in (\ref{gen1.1}) in spherical
coordinates as $d^3 \textbf{r} = r^2 dr \ d\Omega$, where $d\Omega =
\sin\theta \ d\theta \ d\phi$ and $r=[0,\infty)$, $\theta = [0,\pi]$,
and $\phi=[0,2\pi)$, since $f(r)$ has no dependence on $\theta$ or
$\phi$ and since $a_{ijkl}$, $b_{ijkl}$, and $c_{ijkl}$ in
(\ref{pope19}) have no dependence on $r$, the integrals over these
terms in (\ref{gen1.1}) can be considered separately. Using
(\ref{pope18}), the integral in (\ref{gen1.1}) can then be written
as
\begin{eqnarray}\label{gen3.1}
   _{(mp\ldots)}M^{(n)}_{iljk} =
   -\frac{k}{6\pi n!} \left[\int_{0}^\infty r^{n}
  \frac{df}{dr}dr
  \int_\Omega a_{ijkl}\frac{r_m r_p\ldots}{r^n} \ d\Omega
  \right.\\ + \int_{0}^\infty r^{n+1}\frac{d^2 f}{dr^2}dr
  \int_\Omega b_{ijkl}\frac{r_m r_p\ldots}{r^n} \ d\Omega \nonumber
  \\\left.+ \int_{0}^\infty r^{n+2}\frac{d^3 f}{dr^3}dr
  \int_\Omega c_{ijkl}\frac{r_m r_p\ldots}{r^n} \ d\Omega
  \right] \,,\nonumber
\end{eqnarray}
where $k$ in the leading factor is the turbulence kinetic energy.
With the corresponding expression for
$_{(mp\ldots)}M^{(n)}_{jlik}$, (\ref{gen1}) and (\ref{gen3.1})
provide a nonlocal form for the rapid pressure-strain rate
correlation $\Pi^{(r)}_{ij}$ in terms of the longitudinal
correlation $f(r)$.

\subsection{Representing the Longitudinal Correlation $f(r)$}
\label{long_corr} As will be seen later, in (\ref{gen3.1}) the
integrals over $d\Omega$ can be readily evaluated. Moreover, for
$n=0$ the integrals over $dr$ are independent of $f(r)$, and thus
$M^{(0)}_{iljk}$ can be obtained from the general properties
\begin{equation}\label{pope13.1}
  \Lambda = \int_0^\infty f(r) dr\,,\quad f(0)=1\,,\quad
  f(\infty)=0\,.
\end{equation}
However for $n>0$, evaluating the integrals over $dr$ to obtain
$_{(mp\ldots)}M^{(n)}_{jlik}$ requires an explicit form for the
longitudinal correlation function $f(r)$.  We can anticipate,
however, that the precise form may not be of central importance to
our eventual result for $\Pi^{(r)}_{ij}$, since the only role of
$f(r)$ is to
weight the contributions from velocity gradients
$A_{kl}(\textbf{x}+\textbf{r})$
around the local point $\textbf{x}$. It is thus likely that the
integral scale $\Lambda$ in (\ref{pope13.1}) plays the most
essential role, since it determines the size of the region around
$\textbf{x}$ from which nonlocal contributions to the integral for
$\Pi^{(r)}_{ij}$ will be significant. When $r$ is scaled by
$\Lambda$, the precise form of $f(r/\Lambda)$ is likely to be far
less important for most reasonable forms that satisfy the
constraints in (\ref{pope13.1}).

Despite its fundamental significance in turbulence theory, the
form of $f(r)$ for any $r$ and all Reynolds numbers $Re_{\Lambda}$
has yet to be determined even for homogeneous isotropic
turbulence. Perhaps the most widely-accepted representation for
$f(r)$ comes from Kolmogorov's 1941 universal equilibrium
hypotheses. For large values of $Re_{\Lambda} \equiv
k^{1/2}\Lambda / \nu$ and inertial range separations $\lambda_\nu
\ll r\ll \Lambda$, where $\lambda_\nu \sim (\nu^3/\epsilon)^{1/4}$
is the viscous diffusion scale and $\Lambda$ is the integral
length scale in (\ref{pope13.1}), the mean-square velocity
difference is taken to depend solely on $r$ and the turbulent
dissipation rate $\epsilon$, and thus on dimensional grounds must
scale as
\begin{equation}\label{lam2}
  \overline{\left[u'(\textbf{x}+\textbf{r}) -
u'(\textbf{x})\right]^2} \sim \epsilon^{2/3} r^{2/3}\,.
\end{equation}
Expanding the left-hand side of (\ref{lam2}) and using (\ref{frdef})
gives
\begin{equation}\label{lam5}
  \frac{4}{3} k \left[1-f(r)\right] \sim \epsilon^{2/3} r^{2/3}\,.
\end{equation}
Defining the proportionality constant in (\ref{lam5}) as $C_f$ and
rearranging gives the inertial range form of $f(r)$ as
\begin{equation}\label{lam6}
  f(r) = 1- \frac{3}{4} C_f \left[\frac{r}{k^{3/2}/\epsilon}
  \right]^{2/3}\,.
\end{equation}
From Hinze \cite{hinze1975}, a value for $C_f$ can be obtained in
terms of the Kolmogorov constant $K \equiv (8/9\alpha)^{2/3}
\approx 1.7$, where $\alpha\approx 0.405$, as
\begin{equation}\label{lam7}
  C_f = \frac{81}{55}\,\,\Gamma(4/3) K \approx
  2.24\,,
\end{equation}
where we have used $\Gamma(4/3) \approx 0.893$. Expressing
$\Lambda$ in terms of $k$ and $\epsilon$ on dimensional grounds as
\begin{equation}\label{lam1}
  \Lambda = C_\lambda \frac{k^{3/2}}{\epsilon}\,,
\end{equation}
where $C_\lambda$ is a presumably universal constant, then allows
the inertial-range form of $f(r)$ in (\ref{lam6}) to be given as
\begin{equation}\label{irform}
  f(r/\Lambda) = 1- \frac{3}{4} C_f C_\lambda^{2/3} \left(\frac{r}{\Lambda}
  \right)^{2/3}\,.
\end{equation}

\begin{figure}
\centering \setlength{\unitlength}{0.1in}
\includegraphics[width=3.4in]{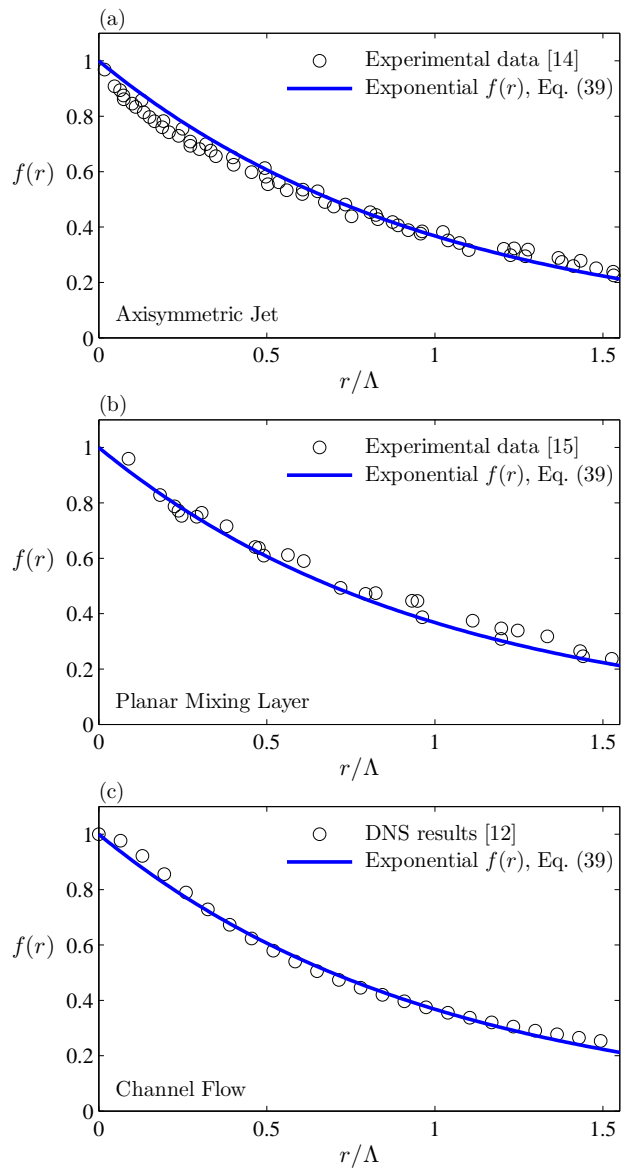}
\caption{Comparison of exponential $f(r)$ in (\ref{gen4}) with
experimental data from axisymmetric turbulent jet
\cite{wygnanski1969} (a) and planar turbulent mixing layer
\cite{wygnanski1970} (b), and with DNS data from turbulent channel
flow at $Re_\tau=650$ \cite{iwamoto2002} (c).}
 \label{frexp}
\end{figure}

\begin{figure}
\centering \setlength{\unitlength}{0.1in}
\includegraphics[width=3.4in]{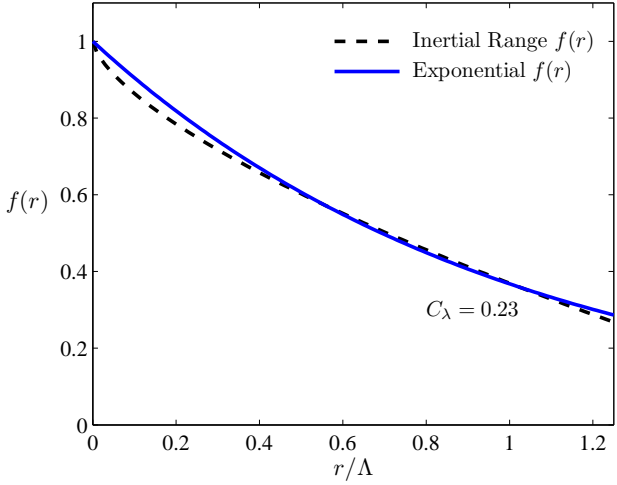}
\caption{Comparison of inertial-range and exponential forms for
$f(r/\Lambda)$ in (\ref{irform}) and (\ref{gen4}), respectively.
Note that $C_\lambda=0.23$ in (\ref{lam1.1}) gives reasonable
agreement between the two forms in the inertial range
$Re_{\Lambda}^{-3/4} \ll (r/\Lambda) \ll 1$}
 \label{fr}
\end{figure}

However the form for $f(r)$ in (\ref{irform}) is valid only for
inertial-range $r$ values, namely $\lambda_\nu\ll r\ll\Lambda$ and
thus for $Re_{\Lambda}^{-3/4} \ll (r/\Lambda) \ll 1$. As a
consequence, this form cannot be used directly to evaluate the
$r$-integrals in (\ref{gen3.1}). However, experimental data from a
wide range of turbulent free shear flows (\textit{e.g.},
\cite{wygnanski1969,wygnanski1970}) and direct numerical
simulation results for wall-bounded turbulent flows
(\textit{e.g.}, \cite{kim1987,iwamoto2002}) show that $f(r)$ can
be reasonably represented by the exponential form
\begin{equation}\label{gen4}
  f(r/\Lambda) = e^{-r/\Lambda}\,,
\end{equation}
as can be seen in Figures \ref{frexp}(a)-(c). Moreover,
$C_\lambda$ in (\ref{lam1}) can be chosen to closely match $f(r)$
in (\ref{gen4}) with the fundamentally-rooted inertial-range form
in (\ref{irform}). Indeed, Figure \ref{fr} shows that with
\begin{equation}\label{lam1.1}
C_\lambda\approx 0.23
\end{equation}
the exponential form in (\ref{gen4}) gives reasonable agreement
with the inertial-range form in (\ref{irform}) up to
$r/\Lambda\approx1$. This exponential form is thus here taken to
represent $f(r)$ in high-$Re_{\Lambda}$ turbulent flows, and will
be used in (\ref{gen3.1}) to obtain an explicit form for the
nonlocal rapid pressure-strain correlation. Since (\ref{gen1})
with (\ref{gen3.1}) is a rigorous formulation for $\Pi^{(r)}_{ij}$
within the central hypothesis on which the present approach is
based, the exponential representation for $f(r)$ is the principal
additional approximation that will be used below in deriving the
present result for the rapid pressure-strain correlation.

While the exponential $f(r)$ appears appropriate for high
$Re_\Lambda$, in the $Re_{\Lambda} \rightarrow 0$ limit the
K\'{a}rm\'{a}n-Haworth equation \cite{karman1938} allows a
solution for $f(r)$.  Batchelor and Townsend \cite{batchelor1948}
showed that when inertial effects can be neglected, this equation
can be solved exactly, giving a Gaussian form for $f(r)$ as
\begin{equation}\label{fr1}
  f(r/ \Lambda) = \exp\left[-\frac{4}{\pi} \left(\frac{r}{\Lambda}\right)^2
  \right]\,.
\end{equation}
Ristorcelli \cite{ristorcelli1998} has proposed a blended form for
$f(r)$ that satisfies various conditions placed on $f(r)$,
including those in (\ref{pope13.1}), while recovering the Gaussian
$f(r)$ in (\ref{fr1}) as $Re_{\Lambda} \rightarrow 0$ and the
exponential $f(r)$ in (\ref{gen4}) as $Re_{\Lambda} \rightarrow
\infty$. It should be possible to use such blended forms for
$f(r)$ to obtain a nonlocal pressure-strain correlation valid for
all Reynolds numbers, following the procedure developed herein. In
the following we obtain the nonlocal pressure-strain correlation
using the high-Reynolds number exponential form in (\ref{gen4}),
which should be accurate for the vast majority of turbulent flow
problems, and then show how this result can be extended to the
low-Reynolds number limit using (\ref{fr1}).

\subsection{Resulting Nonlocal Pressure-Strain Correlation}
Using (\ref{gen4}), it can be shown that the integrals
over $dr$ in (\ref{gen3.1}) give
\begin{subequations}\label{gen5}
\begin{equation}\label{gen5.01}
  \int_0^\infty r^{n} \frac{df}{dr} dr = -
  n!\Lambda^n\,,
\end{equation}
\begin{equation}
  \int_0^\infty r^{n+1} \frac{d^2f}{dr^2}dr =
  (n+1)!\Lambda^n \,,
\end{equation}
\begin{equation}
  \int_0^\infty r^{n+2} \frac{d^3f}{dr^3} dr =
  -(n+2)!\Lambda^n\,.
\end{equation}
\end{subequations}
With these results, (\ref{gen3.1}) is then written as
\begin{eqnarray}\label{gen8}
  _{(mp\ldots)}M^{(n)}_{iljk} =
  k \frac{\Lambda^n}{6\pi} \int_\Omega \left[ \frac{ r_m r_p
  \ldots}{r^n}\right]\hspace{0.9in} \\\cdot\left[a_{ijkl}
  -(n+1)b_{ijkl}+ (n+2)(n+1)c_{ijkl}\right]
  d\Omega\,.\nonumber
\end{eqnarray}
The remaining integrals over $d\Omega$ are all of the form $(r_m
r_p\ldots)/r^n$ and can be solved using the general integral
relations
\begin{subequations}\label{apa3}
\begin{equation}\label{apa3.1}
  \int_\Omega \frac{r_m r_p r_q r_s \ldots}{r^n} d\Omega =
  0\,,\quad  n=\textrm{odd}
  \end{equation}
\begin{eqnarray}\label{apa3.2}
  \int_\Omega \frac{r_m r_p r_q r_s \ldots}{r^n} d\Omega =
  \frac{4\pi}{(n+1)!!}\hspace{1.35in}\\\cdot
  [\delta_{mp}\delta_{qs}\ldots+
  \delta_{mq}\delta_{rs}\ldots +\cdots]\,,\quad
  n=\textrm{even}\nonumber
  \end{eqnarray}
\end{subequations}
where the double factorial is defined as
\begin{equation}\label{gen10}
  n!! \equiv n(n-2)(n-4)\cdots\,,
\end{equation}
with $0!!\equiv1$ and $(-1)!!\equiv1$, and the terms in brackets
on the right-hand side of (\ref{apa3.2}) represent all possible
combinations of delta functions for the indices
$(m,p,q,s,\ldots)$.  For any $n$, there are $(n-1)!!$ such delta
function terms, and each term consists of $(n/2)$ delta functions.

In (\ref{gen8}), for $n=0$ it can be shown using (\ref{apa3.2}) that
$M^{(0)}_{iljk}$ is given by
\begin{eqnarray}\label{gen10.0}
  M^{(0)}_{iljk} = k\frac{2}{15}
  \left(4\delta_{il}\delta_{jk} -
  \delta_{ij} \delta_{kl} -
  \delta_{jl}\delta_{ki}\right)\,.
  \end{eqnarray}
For $n=1$, from (\ref{apa3.1}) $_{m} M^{(1)}_{iljk} = 0$, as
applies to all odd-$n$ cases. For $n=2$, from (\ref{apa3.2})
$_{mp} M^{(2)}_{iljk}$ is given by
\begin{eqnarray}\label{gen10.1}
  _{mp} M^{(2)}_{iljk} =\hspace{2.4in}
  \\k \frac{2 \Lambda^2}{315}[
                        4\delta_{jk}\delta_{il}\delta_{mp}
                        -3\left(\delta_{ij}\delta_{kl}\delta_{mp}+
                        \delta_{jl}\delta_{ik}\delta_{mp}\right)\nonumber
                        \\-24\left(\delta_{il}\delta_{jm}\delta_{kp}+
                        \delta_{il}\delta_{km}\delta_{jp}\right)
  +4(\delta_{ij}\delta_{lm}\delta_{kp}
  +\delta_{ij}\delta_{km}\delta_{lp}\nonumber \\
  +\delta_{jl}\delta_{im}\delta_{kp} +
  \delta_{jl}\delta_{km}\delta_{ip}
  +\delta_{ik}\delta_{lm}\delta_{jp}
  +\delta_{ik}\delta_{jm}\delta_{lp}\nonumber\\
  +\delta_{kl}\delta_{im}\delta_{jp}
  +\delta_{kl}\delta_{jm}\delta_{ip}
  +\delta_{jk}\delta_{im}\delta_{pl}
  +\delta_{jk}\delta_{lm}\delta_{ip})] \nonumber\,.
\end{eqnarray}
Contracting (\ref{gen10.0}) and (\ref{gen10.1}) with $A_{kl}$ and
its derivatives as in (\ref{gen1}) then gives
\begin{equation}\label{gen10.2}
  A_{kl} \left[M^{(0)}_{iljk}+ M^{(0)}_{jlik}\right] =
  \frac{4}{5}k\overline{S}_{ij}\,,
\end{equation}
and
\begin{equation}\label{gen10.3}
  \frac{\partial^2 A_{kl}}{\partial x_m \partial x_p}
   \left[_{mp}M^{(2)}_{iljk}+ _{mp}M^{(2)}_{jlik}\right] =
  \frac{68}{315}k\Lambda^2\nabla^2\overline{S}_{ij}\,,
\end{equation}
where we have used $A_{kk} \equiv 0$. From (\ref{gen10.2}) and
(\ref{gen10.3}), the first two terms in the present formulation
for the pressure-strain correlation in (\ref{gen1}) are thus given
by
\begin{equation}\label{gen10.5}
\frac{1}{k} \Pi_{ij}^{(r)}(\textbf{x}) = \frac{4}{5}\overline{S}_{ij} +
\frac{68}{315}\Lambda^2\nabla^2\overline{S}_{ij}+\cdots\,.
\end{equation}
The first term on the right in (\ref{gen10.5}) is the same as that
in (\ref{ps8}) obtained by Crow \cite{crow1968} assuming spatially
uniform mean velocity gradients. Thus the second term in
(\ref{gen10.5}) is the first-order nonlocal correction accounting
for spatial variations in the mean velocity gradient field.

To obtain the remaining higher-order nonlocal corrections in
(\ref{gen10.5}), it is helpful to contract (\ref{gen8}) with the
derivatives of $A_{kl}$ and again use $A_{kk} \equiv 0$. It is
then readily shown that all terms involving $\delta_{kl}$,
$\delta_{km}$, $\delta_{kp}$, $\ldots$ from the integral over
$d\Omega$ are zero when contracted with the derivatives of
$A_{kl}$, and as a result the coefficients in (\ref{pope19}) can
be simplified as
\begin{subequations}\label{gen12}
\begin{equation}
  a_{ijkl} = 4\delta_{jk} \delta_{il} - \delta_{jl} \delta_{ki} -
  b_{ijkl}\,,
\end{equation}
\begin{equation}
  b_{ijkl} = \delta_{il}\delta_{jk} + b^*_{ijkl}\,,
\end{equation}
\begin{equation}
  b^*_{ijkl} = 3\alpha_{jk} \delta_{il} -
  \delta_{lj} \alpha_{ik} - \delta_{ik} \alpha_{lj} - \delta_{kj}
  \alpha_{il} + 3\beta_{iljk}\,,
\end{equation}
\begin{equation}
  c_{ijkl} = \delta_{il} \alpha_{jk} - \beta_{iljk}\,,
\end{equation}
\end{subequations}
where $b^*_{ijkl}$ has been introduced to simplify the notation.
Using (\ref{gen12}) and contracting (\ref{gen8}) with the
derivatives of $A_{kl}$ we thus obtain
\begin{eqnarray}\label{gen16}
  \frac{\partial^n A_{kl}}{\partial x_m \partial x_p
  \ldots}\left[{_{(mp\ldots)}}M^{(n)}_{iljk}\right] =\hspace{1.5in}\\
  k \Lambda^n\frac{1}{6\pi}
  \left[\frac{\partial^n A_{kl}}
  {\partial x_m \partial x_p \ldots}\right]
   \int_\Omega \left[ \frac{ r_m r_p
  r_q\ldots}{r^n}\right]\nonumber\\
  \cdot[\left(2 - n\right)
  \delta_{il}\delta_{jk} - \delta_{ik} \delta_{jl}
  -\left(n+2\right) b^*_{ijkl} \nonumber \\+
  (n+2)(n+1)c_{ijkl}] d\Omega\,.\nonumber
\end{eqnarray}
From (\ref{apa3.1}) all odd-$n$ terms in (\ref{gen16}) are zero.
For even-$n$, the integrals over $d\Omega$ are readily evaluated
using (\ref{apa3.2}), and it can then be shown that (\ref{gen16})
becomes
\begin{eqnarray}\label{gen32}
  \frac{\partial^n A_{kl}}{\partial x_m \partial x_p
  \ldots}\left[{_{(mp\ldots)}}M^{(n)}_{iljk}\right]
  =\hspace{1.5in}\\k \Lambda^n\frac{2}{3}\left\{ \frac{(n-1)!!}{(n+1)!!}
  \left(\nabla^2\right)^{n/2} \left[
  \left(2 - n\right)
  A_{ji} - A_{ij}\right]\right.\nonumber\\
  -\frac{(n+2)(n-1)!!}{(n+3)!!}
  \left(\nabla^2\right)^{n/2}
  \left[(2-n) A_{ji}-A_{ij}\right]\nonumber\\
  +\frac{(n+2)(n+1)!!}{(n+3)!!}
  \left(\nabla^2\right)^{n/2}
  \left[A_{ji}+A_{ij}\right]
  \nonumber
  \\\left.-\frac{(n+2)(n+4)(n+1)!!}{(n+5)!!}
  \left(\nabla^2\right)^{n/2}
  \left[A_{ji}+A_{ij}\right]\right\}\,.\nonumber
\end{eqnarray}
Adding the corresponding result for $M_{jlik}^{(n)}$ to
(\ref{gen32}) then gives $\Pi^{(r)}_{ij}$ from (\ref{gen1}) as
\begin{equation}\label{gen34}
   \frac{1}{k}\Pi_{ij}^{(r)} = \sum_{n=0,even}^\infty
   \left[C_2^{(n)}\Lambda^n
   \left(\nabla^2\right)^{n/2}\overline{S}_{ij}\right]\,,
\end{equation}
where the coefficients are
\begin{equation}\label{gen33}
  C_2^{(n)} \equiv \frac{4(n^2+2n+9)}
  {3(n+5)(n+3)(n+1)}\,.
\end{equation}
Since the indices in (\ref{gen33}) and (\ref{gen34}) are required
to be even, we can change the index $n$ to $(2n-2)$, where then
$n=1,2,3,\ldots$. This gives the final result for the nonlocal
rapid pressure-strain correlation from the present approach as
\begin{equation}\label{gen10.6}
  \frac{1}{k} \Pi_{ij}^{(r)} = C^{(1)}_2 \overline{S}_{ij} + \sum_{n=2}^\infty
  \left[C_2^{(n)} \Lambda^{2n-2}
   \left(\nabla^2\right)^{n-1}\overline{S}_{ij}\right]\,
\end{equation}
with
\begin{equation}\label{gen10.7}
  C_2^{(n)} \equiv \frac{16n^2-16n+36}{3(2n+3)(4n^2-1)}\,,
\end{equation}
where $\Lambda$ in (\ref{gen10.6}) is from (\ref{lam1}) and
(\ref{lam1.1}). In (\ref{gen10.7}) it may be readily verified that
$C^{(1)}_2 = 4/5$ and $C^{(2)}_2=68/315$, consistent with
(\ref{gen10.5}) and (\ref{ps8}). The first term on the right in
(\ref{gen10.7}) accounts for purely local effects on
$\Pi^{(r)}_{ij}$, while the series term accounts for nonlocal
effects.

The result in (\ref{gen10.6}) and (\ref{gen10.7}) is the first
rigorous formulation for the rapid pressure-strain correlation
$\Pi_{ij}^{(r)}$ that accounts for nonlocal effects due to spatial
variations in the mean velocity gradients. Within the central
hypothesis on which the present approach is based, the principal
approximation used in deriving (\ref{gen10.6}) and (\ref{gen10.7})
is the exponential form of $f(r)$ in (\ref{gen4}) for
high-$Re_\Lambda$ turbulent flows. However, the only effect of
this choice of $f(r)$ is in the resulting coefficients
$C_{2}^{(n)}$ in (\ref{gen10.7}).  All other aspects of
(\ref{gen10.6}) are unaffected by the particular form of $f(r)$,
and instead result directly from the fundamental approach taken
here in solving (\ref{ps4.1}) via Taylor expansion of the mean
velocity gradients $\partial\overline{u}_k / \partial x_l$ to
account for nonlocal effects in $\Pi_{ij}^{(r)}$.

The coefficients $C_{2}^{(n)}$ in (\ref{gen10.7}) from the
exponential representation of $f(r)$ are shown in Figure
\ref{ccoeffs}.  It is apparent that the $n = 1$ term in
(\ref{gen10.6}), which accounts for the purely local contribution
to $\Pi_{ij}^{(r)}$ as verified in (\ref{gen10.5}), is by far the
dominant coefficient. The remaining coefficients for $n = 2, 3, 4,
\ldots$ correspond to the nonlocal contributions to
$\Pi_{ij}^{(r)}$, and can be seen in Fig.\ \ref{ccoeffs} to
decrease only slowly with increasing order $n$. However, while
$C_{2}^{(1)}$ is clearly the dominant coefficient, in
(\ref{gen10.6}) the remaining coefficients are multiplied by
successively higher-order Laplacians of the mean strain rate
field, and thus may produce net contributions to $\Pi_{ij}^{(r)}$
that are comparable to, or possibly even larger than, the local
term due to $n = 1$.

\subsection{Corresponding Coefficients for $Re_\Lambda \rightarrow 0$}
\label{sec_low} 
\begin{figure}
\centering \setlength{\unitlength}{0.1in}
\includegraphics[width=3.4in]{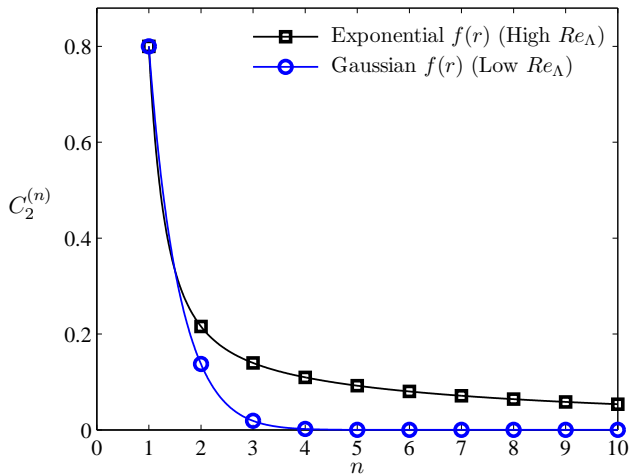}
\caption{Comparison of rapid pressure-strain coefficients $C_2^{(n)}$
from (\ref{gen10.7}) for the $Re_\Lambda \gg 1$
exponential $f(r)$ in (\ref{gen4}), and from (\ref{low10.7}) for the
$Re_\Lambda \rightarrow 0$ Gaussian $f(r)$ in (\ref{fr1}).}
\label{ccoeffs}
\end{figure}

While the coefficients in (\ref{gen10.7}) are
appropriate for $Re_\Lambda \gg 1$, in this Section we use the
exact Gaussian form for $f(r)$ in (\ref{fr1}) that applies in the
$Re_\Lambda \rightarrow 0$ limit to obtain the result for
$\Pi_{ij}^{(r)}$ applicable to low-$Re_\Lambda$ flows, as may
occur in the near-wall region of wall-bounded turbulent flows.
Using this form for $f(r)$, it can be shown that for even-$n$,
which are the only nonzero terms from (\ref{gen8}) due to
(\ref{apa3.1}), the $r$-integrals in (\ref{gen3.1}) are modified
only by multiplying the previous results in (\ref{gen5}) by the
factor $\left[\left(2 / \sqrt{\pi}\right)^{n}(n/2)! / n!\right]$.
The result for $M^{(0)}_{iljk}$ in (\ref{gen10.0}) is independent
of the form of $f(r)$ and thus is unchanged in this limit, but now
$_{mp} M^{(2)}_{iljk}$ in (\ref{gen10.1}) is reduced by the factor
$2/\pi$. With the remaining higher-order terms $_{(mp\ldots)}
M^{(n)}_{iljk}$, it may be readily verified that the result for
$\Pi_{ij}^{(r)}$ in (\ref{gen10.6}) is unchanged in this
low-$Re_\Lambda$ limit, but the coefficients $C_2^{(n)}$ are now
given by
\begin{equation}\label{low10.7}
  C_2^{(n)} = \frac{16n^2-16n+36}{3(2n+3)(4n^2-1)}
  \left[\frac{(n-1)!}{(2n-2)!}
  \left(\frac{4}{\pi}\right)^{n-1}\right]\,,
\end{equation}
where again $n=1,2,3,\ldots$. The effect of the additional factor
in (\ref{low10.7}) relative to (\ref{gen10.7}) is to damp the
higher-order terms in the $Re_\Lambda \gg 1$ coefficients, as
shown in Fig.\ \ref{ccoeffs}.  It is apparent that in this
$Re_\Lambda \rightarrow 0$ limit, only the first nonlocal term ($n
= 2$) in (\ref{gen10.6}) is significant, with all higher-order
coefficients being negligible.  This may introduce significant
simplifications in near-wall modeling, where this limit applies as
$y^+ \rightarrow 0$.

\subsection{Relation to Rotta \cite{rotta1951}}

The present result in (\ref{gen10.6}) with (\ref{gen10.7}) for
$Re_\Lambda \gg 1$ or (\ref{low10.7}) for $Re_\Lambda \rightarrow
0$ is the first nonlocal pressure-strain correlation that
rigorously accounts for the effect of spatial variations in the
mean velocity gradients on the turbulence anisotropy. Previously,
Rotta \cite{rotta1951} derived some of the components of
$M^{(0)}_{iljk}$ and $_{mp} M^{(2)}_{iljk}$, though not enough to
construct even the leading nonlocal term in (\ref{gen10.5}). In
particular, Rotta used an inertial-range form for $f(r)$ similar
to (\ref{lam6}) and the Gaussian form in (\ref{fr1}) to obtain
certain components of $_{mp} M^{(2)}_{iljk}$ in the high- and
low-Reynolds number limits, respectively. Note that Rotta
expressed his results \cite{rotta1951} in terms of the transverse
integral scale $L$ rather than the longitudinal integral scale
$\Lambda$. The two length scales are related by $L=0.5\Lambda$ for
incompressible flows, and this relation can be used directly in
the $Re_\Lambda\rightarrow0$ limit to compare the present result
for $_{mp} M^{(2)}_{iljk}$ with the components obtained by Rotta.
For $Re_\Lambda\gg1$, the differences between the inertial-range
form of $f(r)$ used by Rotta and the exponential form in
(\ref{gen4}) used herein give his $L$ as $L=0.43\Lambda$. Using
these relations it may be verified that the limited components of
$_{mp} M^{(2)}_{iljk}$ given by Rotta are in agreement with the
complete result in (\ref{gen10.1}) for $Re_\Lambda\gg1$, and with
the result for $Re_\Lambda\rightarrow 0$ when the factor of
$2/\pi$ is accounted for as noted in Section \ref{sec_low}.

The agreement with those components of $_{mp}M^{(2)}_{iljk}$
reported by Rotta \cite{rotta1951} provides partial validation of
the present results. However, the present results go much further
by addressing the complete components of $_{(mp\ldots)}
M^{(n)}_{iljk}$ for all $n$, thereby allowing the first complete
formulation of nonlocal effects in the rapid pressure-strain
correlation $\Pi^{(r)}_{ij}$ due to spatial variations in the
mean-flow gradients $\partial \overline{u}_i / \partial x_j$.

\section{Nonlocal Anisotropy Transport Equation}

The present result for nonlocal effects in the
rapid part $\Pi^{(r)}_{ij}$ of the pressure-strain correlation,
given by (\ref{gen10.6}) with the
coefficients $C_2^{(n)}$ in (\ref{gen10.7}) or (\ref{low10.7})
and with $\Lambda$ in (\ref{lam1}) and (\ref{lam1.1}),
can be combined with (\ref{slow1}) for the slow part
$\Pi^{(s)}_{ij}$  to give $\Pi_{ij}$ in (\ref{es4}) as
\begin{eqnarray}\label{aniso1}
  \frac{1}{k} \Pi_{ij} = -C_1 \frac{\epsilon}{k} a_{ij}
  + C^{(1)}_2 \overline{S}_{ij} \hspace{1.5in}\\
  + \sum_{n=2}^{\infty}
  \left[C^{(n)}_2\left(C_\lambda\frac{k^{3/2}}{\epsilon}\right)^{2n-2}
  \left(\nabla^2\right)^{n-1}
  \overline{S}_{ij}\right]\,.\nonumber
\end{eqnarray}
In homogeneous flows, for which prior purely local models for
$\Pi_{ij}$ have been relatively successful, the Laplacians of
$\overline{S}_{ij}$ in (\ref{aniso1}) vanish, and thus the present
nonlocal pressure-strain formulation recovers the local form in
(\ref{ps8}), since $C^{(1)}_2 = 4/5$ in both (\ref{gen10.7}) and
(\ref{low10.7}). For inhomogeneous flows, when (\ref{aniso1}) is
introduced in (\ref{es4}) it gives a new anisotropy transport
equation that accounts for both local and nonlocal effects via the
present fundamental treatment of spatial variations in the mean
velocity gradients in (\ref{ps4.1}).  Note in (\ref{es4}) that the
definition of $P_{ij}$ with $P \equiv P_{nn}/2 \equiv
\overline{u'_i u'_j}\, \overline{S}_{ij}$ gives
\begin{eqnarray}\label{Pij}
  \frac{1}{k} \left[ P_{ij} - \frac{2}{3}P \delta_{ij} \right] \equiv
  -\frac{4}{3}\overline{S_{ij}}+ \left( a_{il} \overline{W}_{lj} -
  \overline{W}_{il}a_{lj} \right)\\
  - \left( a_{il}\overline{S}_{lj} +
  \overline{S}_{il} a_{lj} - \frac{2}{3} a_{nl} \overline{S}_{nl}
  \delta_{ij}\right)\nonumber  \,,
\end{eqnarray}
where the mean-flow rotation rate tensor $\overline{W}_{ij}$
is given by
\begin{equation}
  \overline{W}_{ij} \equiv \frac{1}{2}
  \left( \frac{\partial \overline{u}_i }{\partial x_j}
  - \frac{\partial \overline{u}_j} {\partial x_i}\right)\,.
\end{equation}
From (\ref{Pij}),
the production terms in (\ref{es4}) thus require no additional
closure modeling, while current standard models summarized in
Refs.\ \cite{speziale1991,speziale1998} may be used for the
remaining $\epsilon_{ij}$ and $D_{ij}$ terms.

However, (\ref{aniso1}) does not account for possible
additional anisotropic effects in $\Pi_{ij}^{(r)}$, since the
present nonlocal pressure-strain result in (\ref{gen10.6}) is
based on the central hypothesis that $R_{il}(\textbf{r})$ in
(\ref{gen1.1}) can be represented by its isotropic form in
(\ref{Rijform}). Fundamentally-based approaches for any such
remaining anisotropic effects in (\ref{aniso1}) have yet to be
rigorously formulated, however it has been heuristically argued
($e.g.$, \cite{launder1975,speziale1991a}) that such additional
anisotropy effects may be represented by higher-order tensorial
combinations of $a_{ij}$, $\overline{S}_{ij}$, and
$\overline{W}_{ij}$. The most general of such combinations that
remains linear in $a_{ij}$ is
\begin{eqnarray}\label{Pianiso}
  \frac{1}{k} \Pi_{ij}^{(aniso)} =
  C_3 \left( a_{il}\overline{S}_{lj} +
  \overline{S}_{il} a_{lj} - \frac{2}{3} a_{nl} \overline{S}_{nl}
  \delta_{ij}\right)\\
  + C_4 \left(a_{il} \overline{W}_{lj} -
  \overline{W}_{il}a_{lj}\right)\,,\nonumber
\end{eqnarray}
where the constants $C_3$ and $C_4$ can be chosen to presumably
account for such additional anisotropy effects. In general,
choices for these coefficients vary widely from one model to
another; a summary of various such models is given in Ref.\
\cite{speziale1998}.

When (\ref{aniso1}) is combined with (\ref{Pianiso}),
it provides an anisotropy transport equation that accounts for
both local and nonlocal effects, as well as possible additional
anisotropy effects, in the pressure-strain correlation as
\begin{eqnarray}\label{gen39}
  \frac{Da_{ij}}{Dt} = -\alpha_1
  \frac{\epsilon}{k} a_{ij} +\alpha_2
  \overline{S}_{ij} \hspace{1.7in}\\ + \sum_{n=2}^{\infty}
  \left[C^{(n)}_2\left(C_\lambda\frac{k^{3/2}}{\epsilon}\right)^{2n-2}
  \left(\nabla^2\right)^{n-1}
  \overline{S}_{ij}\right] \nonumber\\
  -\frac{1}{k} \left[ \epsilon_{ij}
  - \frac{2}{3} \epsilon \delta_{ij}\right]\nonumber
  +  \alpha_3\left( a_{il}\overline{S}_{lj} +
  \overline{S}_{il} a_{lj} - \frac{2}{3} a_{nl} \overline{S}_{nl}
  \delta_{ij}\right)\nonumber\\
  - \alpha_4 \left(a_{il} \overline{W}_{lj} -
  \overline{W}_{il}a_{lj}\right)
  +\frac{1}{k} \left[D_{ij} - \left(a_{ij} +
  \frac{2}{3}\delta_{ij}\right) D\right]\,,\nonumber
\end{eqnarray}
where the $C_2^{(n)}$ coefficients are given in (\ref{gen10.7})
or (\ref{low10.7}), and the $\alpha_i$ are defined as
\begin{eqnarray}\label{gen39.1}
  &\alpha_1 = \frac{P}{\epsilon}-1+C_1\,,\quad
  \alpha_2 = C_2^{(1)}-\frac{4}{3}\,,&\\
  &\alpha_3 = C_3-1\,,\quad
  \alpha_4 = C_4-1\,.&\nonumber
  \end{eqnarray}
Values for the constants $C_1$, $C_3$ and $C_4$ in
(\ref{gen39.1}) may be inferred from prior purely local models,
such as the Launder, Reece and Rodi (LRR) \cite{launder1975} or
Speziale, Sarkar and Gatski (SSG) \cite{speziale1991a} models,
which are all based on forms of (\ref{gen39}) without the nonlocal
effects given by the series term. However, optimal values for
these constants may change in the presence of the nonlocal
pressure-strain term in (\ref{gen39.1}).

With respect to the remaining terms in (\ref{gen39}), for high
Reynolds numbers the dissipation tensor $\epsilon_{ij}$ is
concentrated at the smallest scales of the flow, which are assumed
to be isotropic.  Thus, consistent with the central hypothesis on
which the present result for the pressure-strain tensor is
derived, the dissipation is commonly represented by its isotropic
form $\epsilon_{ij}=\frac{2}{3}\epsilon \delta_{ij}$
(\textit{e.g.,} \cite{speziale1998,pope2000}), with the result
that the dissipation term in (\ref{gen39}) vanishes entirely. The
only remaining unclosed terms when (\ref{gen39}) is used with the
ensemble-averaged Navier-Stokes equations are the transport terms
$D_{ij}$ and $D$, and these are typically represented using
gradient-transport hypotheses, with several possible such
formulations summarized in Ref.\ \cite{speziale1998}.

A number of different approaches can be taken for solving
(\ref{gen39}). First, this may be solved as a set of six coupled
partial differential equations, together with the ensemble-averaged
Navier-Stokes equations, to obtain a new nonlocal Reynolds stress
transport closure that improves on existing approaches such as the
LRR and SSG models in strongly inhomogeneous flows. Alternatively,
equilibrium approximations may be used to neglect the $Da_{ij}/Dt$
and $D_{ij}$ terms in (\ref{gen39}) to obtain a new explicit
nonlocal equilibrium stress model for $a_{ij}$, analogous to the
existing local models developed, for example, by Gatski and
Speziale \cite{gatski1993}, Girimaji \cite{girimaji1996}, and
Wallin and Johannson \cite{wallin2000}. Perhaps preferably, a new
explicit nonlocal nonequilibrium stress model for $a_{ij}$ can be
obtained from (\ref{gen39}) following the approach in Ref.\
\cite{hamlington2008a}, by explicitly solving the quasi-linear
form of (\ref{gen39}), namely
\begin{eqnarray}\label{trunc}
  \frac{Da_{ij}}{Dt} = -\alpha_1
  \frac{\epsilon}{k} a_{ij} +\alpha_2
  \overline{S}_{ij} \hspace{1.7in}\\ + \sum_{n=2}^{\infty}
   \left[C^{(n)}_2\left(C_\lambda\frac{k^{3/2}}{\epsilon}\right)^{2n-2}
  \left(\nabla^2\right)^{n-1}
  \overline{S}_{ij}\right]\,.\nonumber
\end{eqnarray}
In so doing it is possible to obtain a new explicit form for the
anisotropy $a_{ij}$ that accounts for both nonlocal and
nonequilibrium effects in turbulent flows.

\section{Conclusions}

A new rigorous and complete formulation for the rapid
pressure-strain correlation, including both local and nonlocal
effects, has been obtained in (\ref{gen10.6}) with (\ref{gen10.7})
for $Re_\Lambda \gg 1$ or (\ref{low10.7}) for $Re_\Lambda
\rightarrow 0$, and with $\Lambda$ in (\ref{lam1}) and
(\ref{lam1.1}). Nonlocal effects are rigorously accounted for
through Taylor expansion of the mean velocity gradients appearing
in the exact integral relation for $\Pi_{ij}^{(r)}$ in
(\ref{ps4.1}). The derivation is based on the central hypothesis
that the nonlocality in $\Pi^{(r)}_{ij}$ is substantially due to
spatial variations in $\partial\overline{u}_k / \partial x_l$ in
(\ref{ps4.1}), and that in order to address this effect all other
factors in (\ref{ps4.1}) can be adequately represented by their
homogeneous isotropic forms. The resulting rapid pressure-strain
correlation in (\ref{gen10.6}) takes the form of an infinite
series of increasing-order Laplacians of the mean strain rate
field $\overline{S}_{ij}(\bf{x})$, with the $n = 1$ term
recovering the classical purely-local form in (\ref{ps8}), and
with the remaining $n \geq 2$ terms accounting for all nonlocal
effects due to spatial variations in the mean-flow velocity
gradients $\partial \overline{u}_k / \partial x_l$.

Aside from the central hypothesis on which the present approach is
based, the sole approximation lies in the need to specify a form
for the longitudinal correlation function $f(r)$.  The particular
specification does not affect the fundamental result in
(\ref{gen10.6}), and serves only to determine the pressure-strain
coefficients $C^{(n)}_2$. For the classical exponential form in
(\ref{gen4}) appropriate for $Re_\Lambda \gg 1$, the corresponding
coefficients are given in (\ref{gen10.7}), while for the exact
Gaussian form in (\ref{fr1}) appropriate for $Re_\Lambda
\rightarrow 0$ the coefficients are given in (\ref{low10.7}). The
integral scale $\Lambda$ in (\ref{gen10.6}) determines the size of
the region around any point over which nonlocal
effects are significant in $\Pi^{(r)}_{ij}$.  In general,
$\Lambda$ can be obtained via (\ref{lam1}), with $C_\lambda$ in
(\ref{lam1.1}) giving good agreement with the inertial-range form
of $f(r)$ in (\ref{lam6}) and (\ref{lam7}) for $Re_\Lambda \gg 1$.

The agreement of the present $n = 1$ term with the purely local
form in (\ref{ps8}) obtained by Crow \cite{crow1968}, and with the
limited components obtained by Rotta \cite{rotta1951} for the
leading $(n = 2)$ nonlocal term, support the validity of the
present derivation. The present results, however, go much further
by accounting for all components $_{(mp\ldots)} M^{(n)}_{iljk}$
for all $n$, which together have allowed the complete form of both
the local and nonlocal parts of the rapid pressure-strain
correlation $\Pi^{(r)}_{ij}$ to be obtained, within the central
hypothesis on which the present approach is based. The present
result thus gives the first rigorous nonlocal form of the rapid
pressure-strain correlation $\Pi^{(r)}_{ij}$ for spatially varying
mean velocity gradients in turbulent flows.

Using the present result for $\Pi_{ij}^{(r)}$ in (\ref{gen10.6})
with (\ref{lam1}) and (\ref{lam1.1}) and
with (\ref{gen10.7}) or (\ref{low10.7}), a nonlocal transport
equation for the turbulence anisotropy has been obtained in
(\ref{gen39}) and (\ref{gen39.1}). The resulting nonlocal
anisotropy equation can be solved by any number of standard
methods, including full Reynolds stress transport closure
approaches, algebraic stress approaches, or the nonequilibrium
anisotropy approach outlined in Ref.\ \cite{hamlington2008a} based
on (\ref{trunc}). This nonlocal anisotropy equation should give
significantly greater accuracy in simulations of inhomogeneous
turbulent flows, including free shear and wall-bounded flows,
where strongly nonuniform mean flow properties and significant
large scale structures will introduce substantial nonlocal effects
in the turbulence anisotropy.

\acknowledgments This work was supported, in part, by the Air
Force Research Laboratory (AFRL) through the Michigan-AFRL-Boeing
Collaborative Center for Aeronautical Sciences (MAB-CCAS) under
Award No.\ FA8650-06-2-3625.

\bibliographystyle{unsrt}

\end{document}